\documentclass{caosp305}

\usepackage{graphicx}

\usepackage{natbib}
\articleNo{123}
\pubyear{2005}
\volume{35}
\volnumber{3}
\firstpage{1}
\received{May 1, 2005}
\accepted{August 28, 2005}

\def\BibTeX{{\rm B\kern-.05em{\sc i\kern-.025em b}\kern-.08em
             T\kern-.1667em\lower.7ex\hbox{E}\kern-.125emX}}

\begin{document}

\hauthor{H.\,Korhonen}

\title{Magnetic fields of cool giant and supergiant stars: models versus observations}


%
\author{H.\,Korhonen \inst{1} 
       }

\institute{
  Dark Cosmology Centre, Niels Bohr Institute, University of Copenhagen\\
  DK-2100 Copenhagen, Denmark
          }

\date{March 8, 2003}

\maketitle

\begin{abstract}
The recent years have brought great advances in our knowledge of magnetic fields in cool giant and supergiant stars. For example, starspots have been directly imaged on the surface of an active giant star using optical interferometry, and magnetic fields have been detected in numerous slowly rotating giants and even on supergiants. Here, I review what is currently known of the magnetism in cool giant and supergiant stars, and discuss the origin of these fields and what is theoretically known about them.
\keywords{Stars: atmospheres -- Stars: evolution -- Stars: late-type -- Stars: magnetic field}
\end{abstract}

\section{Introduction}

\citet{2014IAUS..302..350K} presented a review of the magnetism in cool giant and supergiant stars. The current review concentrates on the advances that have been made since then. The emphasis of this review is on actual magnetic field measurements, but also some discussion is included on imaging starspots and other interesting developments in the field.

\section{Imaging starspots}

Over the years starspot locations have been imaged from photometry and high resolution spectroscopy. The modelling of the photometric light-curves gives only accurate information on the longitudinal location of the spots \citep[see, e.g.,][]{2010AN....331..250V}. Using high resolution, high signal-to-noise spectroscopic observations and inversion techniques \citep[Doppler imaging, see, e.g.,][]{1983PASP...95..565V} gives information on both the spot longitude and latitude \citep[e.g.,][]{2007A&A...476..881K}. In the following two recent advances on imaging surfaces of giant stars are discussed.

\subsection{Interferometric imaging}

During the last decade significant advances have been made in using long-baseline interferometric imaging. Nowadays, there are facilities that can combine the light from six telescopes and produce high fidelity images. The latest break through in interferometric imaging comes from imaging starspots on the surface of RS\,CVn-type active K giant $\zeta$\,Andromedae \citep{2016Natur.533..217R}.

\citet{2016Natur.533..217R} imaged $\zeta$\,And using the Michigan Infrared Combiner (MIRC) at Centre for High Angular Resolution Astronomy (CHARA) on Mt.\,Wilson. MIRC combines the light from all the six telescopes in the CHARA-array allowing for high fidelity imaging. The images obtained by \citet{2016Natur.533..217R} at two different epochs, 2011 and 2013, show persistent polar spot and changing spots at lower latitudes. The images also revealed, for the first time,  reliable starspot hemispheres (north vs south). In a very recent paper \citet{2017ApJ...849..120R} also critically discusses the pros and cons of different techniques for imaging stellar surface structures: light-curve inversions, Doppler imaging, and interferometric imaging.

\subsection{Magnetic field maps from all four Stokes parameters}

In another recent development \citet{2015ApJ...805..169R} used full four Stokes parameters in the magnetic field mapping of a RS\,CVn-type binary star II\,Peg. Typically only circular polarisation (Stokes V) is used in magnetic field mapping. The linear polarisation (Stokes Q and U) signals from starspot are weak, and linear polarisation also requires sophisticated radiative transfer modelling for interpreting the results.

\begin{figure}
\centerline{\includegraphics[width=0.95\textwidth,clip=]{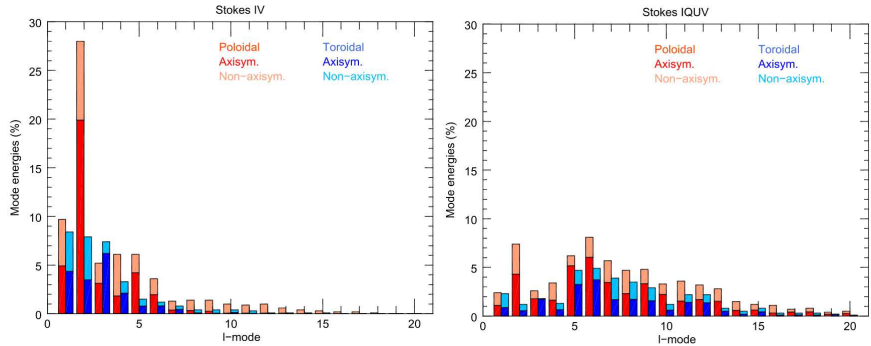}}
\caption{Distribution of the magnetic energy in different harmonic modes for the map obtained only from circular polarisation (Stokes IV) and also taking into account linear polarisation (Stokes IVQU). \citep[From][]{2015ApJ...805..169R}}
\label{Rosen}
\end{figure}

\citet{2015ApJ...805..169R} show the difference between maps obtained only from circular polarisation (Stokes I and V) and using full four Stokes parameters (Stokes I, V, Q, and U). Their results clearly show that the strength of some of the surface features increases significantly (doubles or even quadrupled) when linear polarization is taken into account. At the same time, the total magnetic energy of the reconstructed field becomes significantly higher, and the over-all field complexity increases. This is well illustrated in Fig. \ref{Rosen}, which shows the magnetic field energy in different spherical harmonic modes for IV and IVQU maps.

The study of \citet{2015ApJ...805..169R} underlines the importance of using all the four Stokes parameters, if possible, when doing Zeeman Doppler imaging. This result has been shown before for the hot stars \citep[e.g.,][]{2000MNRAS.313..851W}, but is now also convincingly expanded to cool stars.

\section{Magnetic field measurements in cool giant stars}

The following sections discuss the recent measurements of magnetic fields in stars when they evolve away from the main sequence towards the asymptotic giant branch (AGB). AGB stars are not in the scope of this review, as they were discussed in another review talk (Vlemmings 2017, these proceedings).

\subsection{Hertzsprung gap stars}

When a star leaves the main sequence, it will first cross the so-called Hertzsprung gap. An area where there are only a few stars due to the fast evolution in this phase.

Magnetic field measurements, and also surface magnetic field maps, have been obtained of few stars in the Hertzsprung gap. For example, \citet{2017A&A...599A..72T} looked at 37\,Com, a 6.5\,M$_\odot$ Hertzsprung gap star, and recover magnetic field configuration that has mainly poloidal geometry. \citet{2016A&A...591A..57B} compared the behaviour of two Hertzsprung gap giants, OU\,And and 31\,Com, which have very similar masses but different rotation period. OU\,And has rotation period of 24.2 days and mass of 2.7 M$_\odot$, whereas 31\,Com has rotation period of 6.8 days and mass of 2.85 M$_\odot$. Magnetic fields are detected in both stars. Surprisingly, the faster rotating 31\,Com has a weaker field with complex topology, and the slower rotating OU\,And has a stronger field and largely dipolar topology. \citet{2016A&A...591A..57B} concluded that the field in 31\,Com was most likely dynamo created, and that OU\,And was possibly a descendant of a magnetic Ap star.

\subsection{Red giant stars}

In the recent years magnetic field measurements in numerous red giant stars have been obtained. \citet{2015A&A...574A..90A} studied magnetic fields in a sample of 48 evolved cool stars. From the target stars 24 were known to show signs of magnetic activity. The study found definite detection of magnetic field in 29 targets, and Zeeman signatures were found in all but one of the 24 active red giants. The additional six stars showing magnetic fields were bright red giants. When comparing the activity index {\it S} to the magnetic field detection, one can see that the sample of \citet{2015A&A...574A..90A} has detections mostly on stars with higher activity levels. 

However, also really slowly rotating red giants can exhibit weak, but detectable, magnetic fields. Weak, sub-Gauss, magnetic fields have been detected for example in Pollux \citep[K0 III, period $\sim$100-500d,][]{2015A&A...574A..90A}, and Arcturus \citep[K1.5 III, $\sim$730d,][]{2011A&A...529A.100S}. \citet{2015A&A...574A..90A} detect weak fields also on Aldebaran, Alphard, and $\eta$\,Psc. \citet{2015A&A...574A..90A} suggested that a solar-like $\alpha\Omega$-dynamo driven by convection and differential rotation is operating in these stars.

Theoretical calculations by \citet{2017A&A...605A.102C} show that Dynamo processes might be favoured in the stellar convective envelope at two specific moments at the later stages of stellar evolution: during the first dredge-up, and during central helium burning in the helium-burning phase and early-AGB. This is nicely supported by the results of \citet{2015A&A...574A..90A}.

\subsection{The special case of EK Eri}

EK\,Eridani is a special case among magnetic cool giants. Its highest measured longitudinal magnetic field is $\sim$100G \citep{2015A&A...574A..90A}, but its rotation period is surprisingly long, $\sim$300 days \citep{1999A&A...343..175S}.

\begin{figure}
\centerline{\includegraphics[width=0.75\textwidth,clip=]{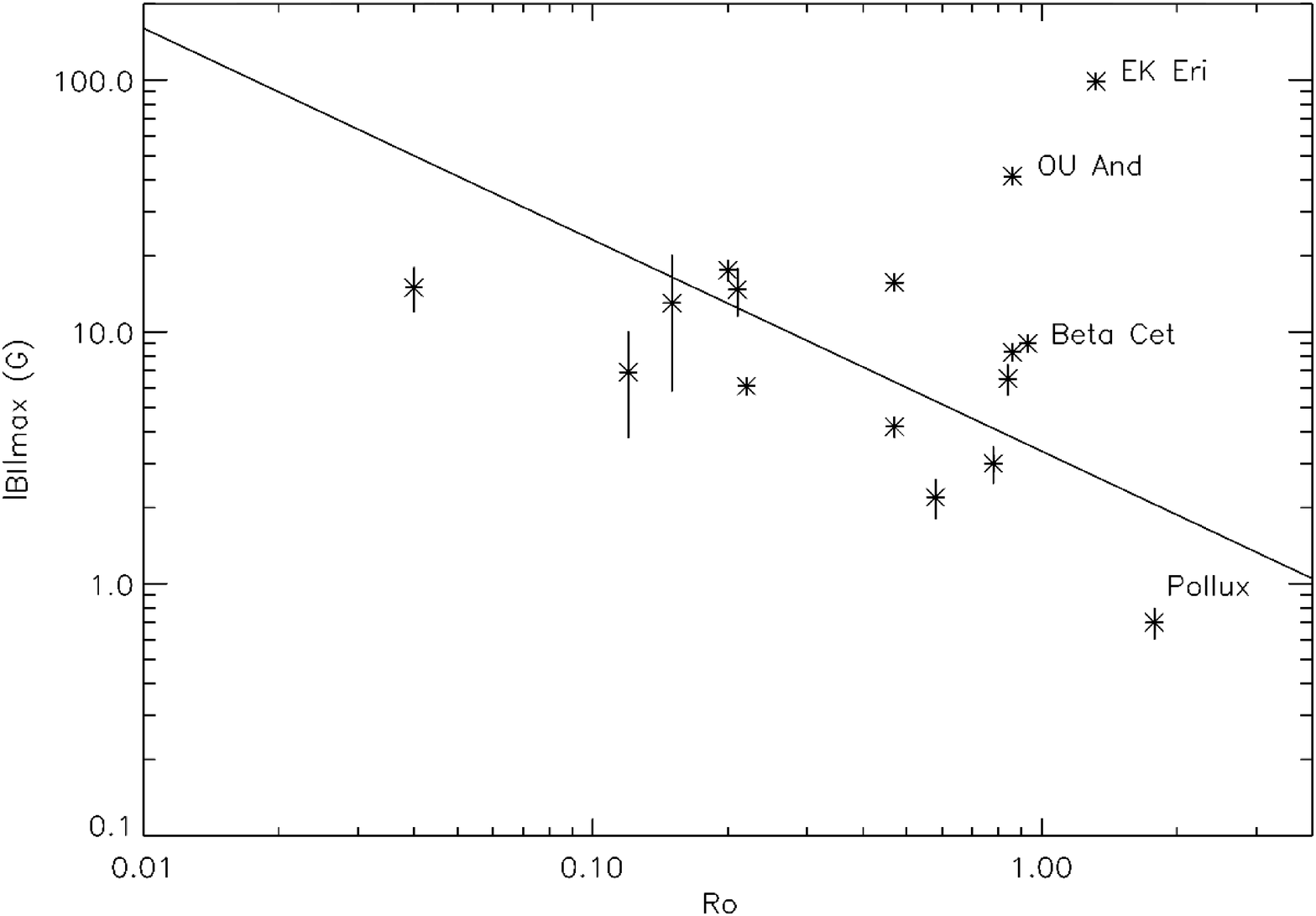}}
\caption{Correlating Rosby number and longitudinal magnetic field strength. \citep[From][]{2015A&A...574A..90A} }
\label{Ro_B}
\end{figure}

Stars with dynamo created magnetic fields tend to scatter around a line, if one plots their Rosby number (rotation period divided by the convective turnover time) against the longitudinal magnetic field strength. In Fig.\,\ref{Ro_B} stars in \citet{2015A&A...574A..90A} sample are plotted. The location of EK\,Eri this diagram implies that its magnetic field is most likely not dynamo created. Therefore, it is most likely a descendant of a strongly magnetic Ap star \citep{2015A&A...574A..90A}.

However, the formation of the convective outer layer during the transition from main sequence to red giant phase would initiate dynamo-created field. This should destroy the earlier fossil field with very short time scales (Arlt \& Bonnano, private communication). The location of EK\,Eri at the base of the red giant branch would imply that the possible fossil field should have already been destroyed. Another star that is an outlier in Fig.\,\ref{Ro_B} is OU\,And. But, as discussed earlier, this star is still in the Hertzsprung gap, and therefore could have still maintained its fossil field.

Also $\beta$\,Cet has been proposed to be a descendant of Ap stars \citep{2013A&A...556A..43T}, but its location on the Rosby number vs magnetic field strength diagram, as seen in Fig.\,\ref{Ro_B}), implies that the magnetic field is most likely dynamo created.

\section{Cool supergiants}

\citet{2010MNRAS.408.2290G} carried out an extensive study of magnetic fields in supergiants. Their sample of 33 supergiants spanned the spectral types M1.5Iab to A0Ib. They detected fields in nine of the sample stars, spanning spectral types F--K.

The first M supergiant star with detected magnetic fields was Betelgeuse \citep{2010A&A...516L...2A}, and the fields have been proposed to concentrate in the sinking components of the convective flows \citep{2013LNP...857..231P}. Recently weak magnetic fields have also been detected in two other red supergiants: CE\,Tau and $\mu$\,Cep \citep{2017A&A...603A.129T}. It has been postulated that the fields in supergiants can be created by local dynamo operating in the giant convective cells \citep{2004A&A...423.1101D}.

\section{Other interesting developments}

For this last section I have selected two interesting ideas from recent papers dealing with magnetic fields in evolved stars.

\subsection{Dynamo enhancement due to engulfment of planets?}

If a star engulf planets during the later stages of its evolution, it will also acquire the angular momentum of those planets. This, on the other hand, increases the rotation rate of the star and enhances dynamo operation

\citet{2016A&A...593L..15P} show that the engulfment of a 15 M$_{J}$ planet produces a dynamo triggered magnetic field stronger than 10 G for gravities between 2.5 and 1.9. They also show that for reasonable magnetic braking laws, the high rotation rate induced by a planet engulfment may be maintained sufficiently long to be observable.

\subsection{The mysterious case of red giant oscillations}

20\% of red giants have suppressed l=1 modes. \citet{2015Sci...350..423F} show that the suppression can be explained if waves entering the stellar core are prevented from returning to the envelope. \citet{2016Natur.529..364S} saw no suppression in red giants below 1.1 M$_{\odot}$, and the incidence of magnetic suppression increases with mass, with red giants above 1.6 M$_{\odot}$ showing a suppression rate of 50\% to 60\%. Interestingly, this is the mass range where main sequence stars have convective cores. These results could imply magentic field creation in the convective cores, as predicted by, e.g., \citet{2016ApJ...829...92A}

\section{Conclusions}

As is evident from this review, recent years have shown many new developments in observing magnetic fields in cool giant and supergiant stars. Our observations show that magnetic fields are indeed present in basically all the stellar evolutionary stages -- at least weak fields.

The origin of these magnetic fields are roughly from two different origins: $\alpha\Omega$-type dynamo process operating in the convective envelope, and fossil fields. It has also been postulated that dynamos can operate in giant convective shells of super giant stars. And there is also increasing evidence that higher mass stars can have dynamos operating also in their convective cores.

\acknowledgements
HK acknowledges financial support from the conference organisers and the Danish Augustinus foundation which made her participation in this interesting conference possible.

\bibliography{Magnetic17_Korhonen}

\end{document}